\documentclass{optica-article}

\journal{opticajournal} 

\articletype{Research Article}

\usepackage{lineno}
\linenumbers 

\begin{document}
\nolinenumbers
\title{Off-axis holographic imaging with undetected light}

\author{Josué R. León-Torres,\authormark{1, 2, 3*} Filip Krajinić,\authormark{4, 5} Mohit Kumar,\authormark{1} Marta Gilaberte Basset, \authormark{1, 2} Frank Setzpfandt, \authormark{1, 2} Valerio Flavio Gili,\authormark{2} Branislav Jelenković,\authormark{4, 5} and Markus Gräfe\authormark{2, 6}}

\address{\authormark{1}Abbe Center of Photonics, Friedrich Schiller University Jena, Albert-Einstein-Straße 6, 07745 Jena, Germany\\
\authormark{2}Fraunhofer Institute for Applied Optics and Precision Engineering IOF, Albert-Einstein-Straße 7, 07745 Jena, Germany\\

\authormark{3} Cluster of Excellence Balance of the Microverse, Friedrich Schiller University Jena, Jena, Germany\\

\authormark{4} University of Belgrade, School of Electrical Engineering, Bulevar kralja Aleksandra 73, 11120 Belgrade, Serbia\\
\authormark{5} University of Belgrade, Institute of Physics Belgrade, Photonics Center, Pregrevica 118, 11080 Zemun, Belgrade, Serbia\\
\authormark{6} Institute of Applied Physics, Technical University of Darmstadt, Schloßgartenstraße 7, 64289 Darmstadt, Germany}

\email{\authormark{*}josue.ricardo.leon.torres@iof.fraunhofer.de} 


\begin{abstract*} 
Quantum imaging with undetected light (QIUL) can retrieve amplitude and phase information of an object by exploiting the quantum correlations of photon-pairs generated through spontaneous parametric down conversion (SPDC), where the illumination and detection can be carried at very distinct wavelength ranges. This fact allows to benefit from a mature detection technology in the visible spectral range, while probing the object at a more exotic wavelength. Here we experimentally implement a QIUL approach with Fourier off-axis holography in a hybrid-type induced-coherence non-linear interferometer. Our approach reconstructs the amplitude and phase information of an object with a single shot in a wide-field configuration, being an alternative in front of techniques that require multiple acquisition frames, such as phase-shifting holography.

\end{abstract*}

\section{Introduction}

There are many applications in which holographic imaging techniques are demanded, ranging from manufacturing, in-line quality control \cite{Fratz2021, Besaga2019} to medical research and diagnosis \cite{Erickson1996}, with particular attention on the study of cells and tissues, quantitative phase imaging has come up as a powerful method \cite{Popescu, Nguyen}. 

Over the years, different imaging schemes have been developed to study transparent organisms, such as phase contrast microscopy, first proposed by Zernike in 1932 \cite{Zernike} and numerous techniques based on holography, first introduced by Gabor in 1948 \cite{Gabor, Gabor1}. 

The quantum revolution contributed to develop holographic imaging schemes that exploit the quantum nature of light in order to overcome classical limitations, for instance to improve the phase sensitivity \cite{Boyd, Silberhorn, Genovese}, to study fundamental concepts such as amplitude and phase of a single photon \cite{Teich, Wasilewski, Kundu, Denis, Tekadath, Zia} and to allow different wavelengths for detection and sample illumination \cite{Fuenzalida2, Andras, Sebastian, Haase2023}. 

In addition, non-linear interferometry has proven to be a useful tool in a multitude of quantum optics related fields such as quantum communication, imaging, sensing, and information theory \cite{Ou, Yang, Fuenzalida2, Kim2024}. Non-linear interferometers integrate elements such as parametric amplifiers, and second order non-linear crystals to generate photon-pairs via SPDC. Furthermore, they allow a controlled splitting and mixing of probability amplitudes to precisely estimate phase changes of the optical field \cite{Ou, Chekhova2016}.

QIUL combines elements of non-linear interferometry and quantum optics, in which a laser pumps a non-linear crystal to generate spatially correlated photon-pairs, historically named as signal and idler, via SPDC \cite{Kutas2022, Chekhova2016, Walborn2010}. The wavelengths of the signal and idler beams can be easily tuned by controlling the pump and crystal parameters. The scheme illuminates the sample with the idler beam at a particular wavelength which can be very distinct from the detected signal beam, particularly taking advantage of the well developed silicon-based technology \cite{Marta, Kviatkovsky2020, Kutas2022}.

The present work is a step forward from the first implementation of holography with undetected light based on digital phase-shifting holography (DPSH) \cite{Sebastian}, which needs to take a minimum of three images to reconstruct the object information \cite{Yamaguchi1997}. In contrast, our approach based on Fourier off-axis holography (OAH) \cite{Gabor, Kim, Sanchez} takes a single image and retrieves amplitude and phase information, having no necessity of utilizing a piezo stage to have control of the overall interferometer phase, and reducing the cost and complexity of the system. 



\section{Non-linear interferometer for quantum OAH}

\subsection{In-line and off-axis classical holography}

Since its first appearance, holography has become a widely used method in a countless number of fields, ranging from microscopy, interferometry to security and data storage \cite{Kim, Popescu, Nguyen}. In this section we will discuss the two most common approaches for classical holography known as in-line and OAH. 

In classical holography, the amplitude of a laser beam is split in half through a 50:50 beam splitter (BS) placed at the center of a Michelson-Morley interferometer, generating the object and reference beams\cite{Sanchez, Verrier}, see Fig 1 a). The object beam interacts with the sample and is back reflected by a mirror placed at the end of the arm. The two beams meet again at the BS and are redirected to the camera plane to form the hologram. In-line holography extracts the object information when the object and reference beams are spatially overlapped, their relative angle is zero. In contrast, in OAH a relative angle between object and reference beams is introduced, for example by means of the tip-tilt of a mirror, adding a linear phase to the reference beam \cite{Cuche, Sanchez, Verrier, Kim}.

In-line holography is usually implemented through DPSH, which requires at least three measurements to retrieve the amplitude and phase information carried by the optical field \cite{Yamaguchi1997}. One of the drawbacks of this technique is the limitation to study samples whose dynamics changes over timescales comparable or shorter than the typical duration of one of the three required measurements. OAH rises up as an alternative that requires only a single measurement to reconstruct the amplitude and phase carried by the field \cite{Kim, Cuche, Sanchez}, therefore allowing to study dynamic behavior within a shorter time window relative to DPSH. These two competing methods differentiate from each other on how the interfering intensity components at the camera plane are distributed in the Fourier space (FS). Figures 1 b) and c) showcase the relative angle between object and reference beams and their influence on how their intensity components spread in the FS.

Let us introduce the intensity distribution of the OAH recorded at the camera plane \cite{Cuche, Verrier, Sanchez}. 
\begin{equation}
    I(\boldsymbol{r})=|E_{O}(\boldsymbol{r})+E_{R}(\boldsymbol{r})|^{2}=|E_{O}(\boldsymbol{r})|^{2} +|E_{R}(\boldsymbol{r})|^{2}+E_{O}^{*}(\boldsymbol{r})E_{R}(\boldsymbol{r})+E_{R}^{*}(\boldsymbol{r})E_{O}(\boldsymbol{r}),
\end{equation}

its components are the four terms that surge out of the squared coherent sum of the object and reference fields. The first two terms correspond to the amplitude squared of object $E_{O}(\boldsymbol{r})$ and reference $E_{R}(\boldsymbol{r})$ fields. The third and fourth are the interference terms, which depend on the relative phase between them, the asterisk denotes complex conjugation. 

The Fourier transform (FT) of Eq. (1) leads to

\begin{equation}
    \Tilde{I}(\boldsymbol{q})=DC(\boldsymbol{q})+ \Tilde{E}_{O}^{*}(\boldsymbol{q})\otimes\delta(\boldsymbol{q}-\boldsymbol{k})+\Tilde{E}_{O}(\boldsymbol{q})\otimes\delta(\boldsymbol{q}+\boldsymbol{k}),
\end{equation}

where the tilde means FT with respect to position $\boldsymbol{r}$, $\otimes$ denotes linear convolution, and $\boldsymbol{q}$ spatial frequency. The first term is the auto-correlation term or the FT of the first two terms of Eq. (1) and the second and third are the direct and conjugate terms, respectively. These last two terms exhibit a displacement proportional to the wavevector $\boldsymbol{k}$, usually carried by the reference beam, $E_{R}(\boldsymbol{r})=A_{R} e^{i\boldsymbol{k}\cdot\boldsymbol{r}}$, where $A_{R}$ is its amplitude. They are often named as the $+1$ and $-1$ diffraction orders, respectively. In DPSH, due to the full spatial overlap between the object and reference beam, this splitting does not take place, as shown in Fig. 1 b). The wavevector $\boldsymbol{k}$ is introduced in OAH with the sole purpose of splitting the first order terms with respect to the zero order in the FS, shown in Fig. 1 c), such that the direct term can be filtered and isolated to obtain a complex-value image by doing an inverse Fast Fourier Transform (IFFT). The modulus corresponds to the amplitude $A_{O}(\boldsymbol{r})$ and the argument value corresponds to the phase $\phi_{O}(\boldsymbol{r})$ of the object field \cite{Kundu, Sanchez, Verrier}. The wavevector $\boldsymbol{k}$ is proportional to the relative angle between the object and reference beams, this is induced by mechanical means such as tip-tilt of mirrors. 




The features of quantum OAH are governed by the same principles that hold classically, such as the larger the relative angle the higher the density of fringes in the hologram, and depending on how the beam is tilted the $+1$ and $-1$ terms open up along a vertical, horizontal, diagonal or anti-diagonal axis. A detailed analysis is presented in Sec. 2.2.

\begin{figure}[ht!]
\centering\includegraphics[width=13.2cm]{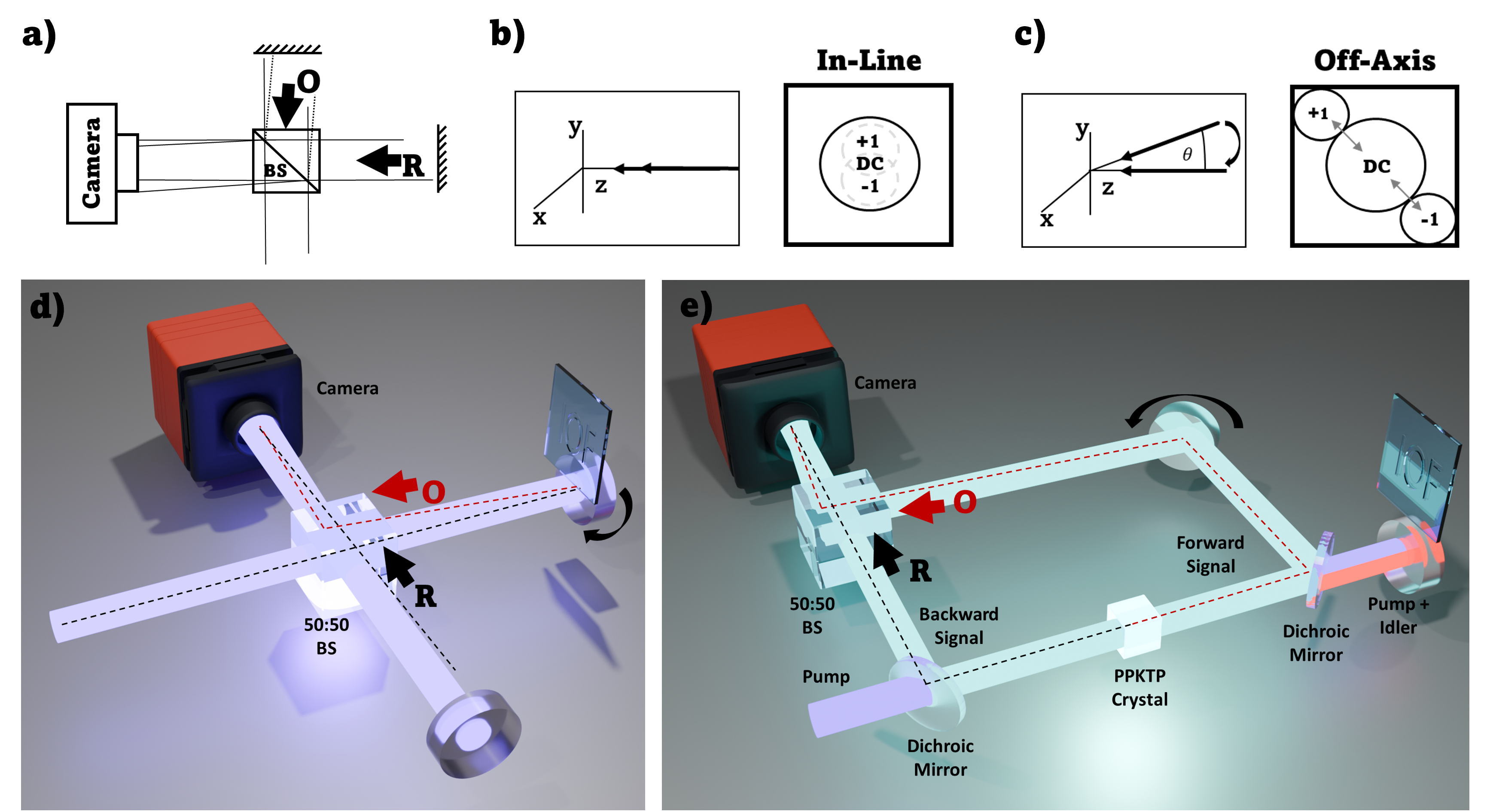}
\caption{Comparison between in-line and OAH (upper row) and classical and quantum OAH (lower row). a) A coherent beam is sent to a Michelson-Morley interforemeter. The beam that hits the object is tilted in comparison to the reference beam, allowing the $+1$, DC, and $-1$ terms to split in the FS as seen through Figs. b) and c). d) Classical implementation of OAH. e) Quantum implementation of OAH, the purple beam pumps the non-linear crystal in forward and backward direction, generating signal and idler beams via SPDC in both ways. A relative angle is introduced between object and reference beams, they recombine at the BS and interfere at the camera plane generating a hologram.}
\end{figure}

\subsection{Classical vs quantum off-axis holography with undetected light}

The basic principles of classical holography also applies to its quantum counterpart, in which the hologram is formed by the interference of the object and reference beams \cite{Gabor}. What differentiates the quantum from the classical implementation is the nature of these beams. In classical holography, the object and reference beams originate by splitting the amplitude of a coherent laser through a beam spliter \cite{Verrier, Sanchez}, as shown in Fig. 1 a) and d), meaning that the two beams exist simultaneously and overlap spatially and temporally at the camera plane. \\
In contrast, in quantum off-axis holography with undetected light (QOAHUL), simplified sketch shown in Fig. 1 e), a coherent beam depicted as the purple beam pumps a non-linear crystal bidirectionally, generating photon-pairs in both directions via SPDC \cite{Marta, Kviatkovsky2020}. We call the forward generated signal beam the object beam (cyan beam with red dashed line), its idler twin beam interacts with the object and is back reflected to the crystal. The backward generated signal beam is known as the reference beam (cyan beam with black dashed line), its idler twin beam overlaps spatially with the forward generated idler at the crystal plane to induce coherence on their signal beams \cite{Mandel, Mandel2}, allowing them to interfere at the camera plane. 

In the low-gain regime, at any given time, only one pair of photons is being generated, therefore the hologram observed at the camera plane is the result of the interference of the probability amplitudes associated to the generation of the object and reference beams, which never interacted with the object \cite{Lemos2014, Kundu}. 

In QOAHUL the object and reference beams are capable of interfering as a consequence of the coherence induced by the spatial and temporal overlap of their twin idler beams within the coherence time of the pump beam \cite{Mandel, Mandel2, Molker}. This overlap ensures that there is no knowledge of the which-path information, in other words, the two idler beams become indistinguishable and the observer is incapable of knowing whether the signal photon that strikes the camera plane was generated in the forward or backward path resulting in a single-photon interference pattern \cite{Young, Mandel, Mandel2, Lahiri}. The reconstruction of the object information is also attributed to the spatial correlations shared by the photon-pair, namely momentum correlations which are the result of momentum conservation in the SPDC process \cite{Lemos2014, Walborn2010}. Furthermore, the undetected light scheme grants the opportunity to work in a very distinct wavelength range for the illumination and detection parts, taking advantage of a mature silicon based detection technology at the visible wavelength range \cite{Kutas2022, Kviatkovsky2020, Marta}. 

\subsection{Experimental setup}

Figure 2 shows a sketch of the implemented QOAHUL setup. A periodically poled potassium titanyl phosphate (ppKTP) non-linear crystal of 2 mm length is pumped with a 20 mW CW laser at 405 nm, signal (910 nm) and idler (730 nm) photons are produced by SPDC. The pump beam depicted in purple, travels in forward and backward directions through the ppKTP crystal generating two probability amplitudes associated to the photon-pair generation events in both directions \cite{Marta, Lemos2014}. Both events are considered to be equally probable. When the photon-pair is generated in the forward path, the signal beam depicted in cyan and idler beam depicted in red are split by off the shelf dichroic mirror, DM-1. The idler and pump beams are split by DM-2, letting the idler beam to propagate through a Michelson-Morley like interferometer. At the end of the interferometer arm the idler beam traverses the object and is back reflected to the crystal by a mirror where it spatially overlaps with the backward generated idler beam. A proper alignment of these beams induces coherence onto their signal twins allowing them to interfere at the camera plane \cite{Mandel, Mandel2}

The forward and backward generated signal beams propagate through the equivalent of the lower and upper arms of a Mach-Zender interferometer (MZI). By tip-tilting any of the mirrors right before the BS a relative angle between object and reference beams is introduced as discussed in section 2.1 \cite{Cuche, Sanchez, Verrier}. The forward and backward signal beams meet at the 50:50 BS and are redirected to the camera plane. The MZI offers a better control of the optical paths between the two beams providing a large relative angle, which can optimize the imaging capabilities of the system \cite{Sanchez, Verrier}.
The signal beam allows the image reconstruction despite never interacting with the object as a result of the induced coherence and the spatial correlations between signal and idler beams \cite{Mandel, Walborn2010}. 

\begin{figure}[ht!]
\centering\includegraphics[width=13.2cm]{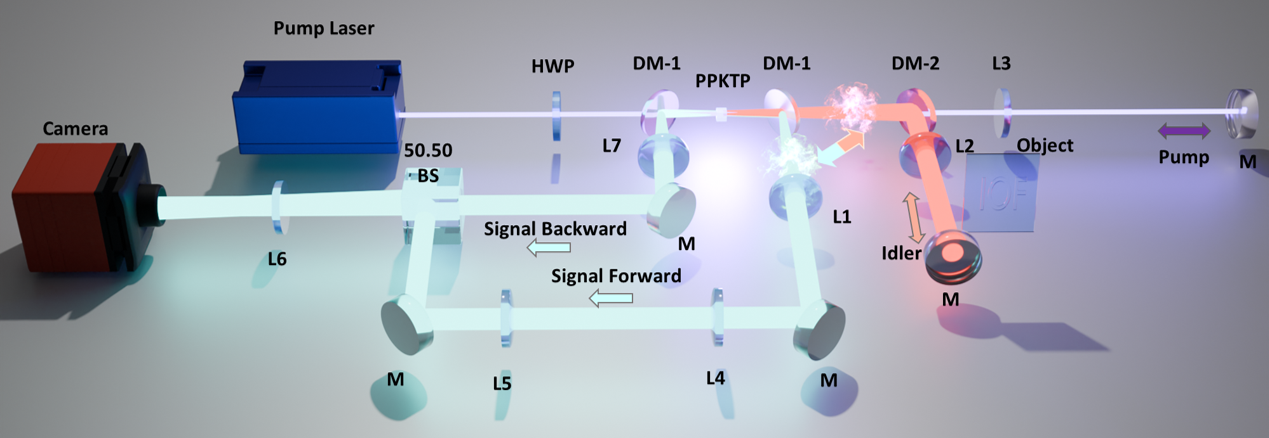}
\caption{Experimental setup: a 2 mm long ppKTP non-linear crystal is pumped by a 405 nm CW laser, it generates a photon pair in the forward and backward direction, signal and idler beams at 910 nm (cyan beam) and  730 nm (red beam), respectively. Although the signal beam never interacts with the object (located at the idler path) is utilized for the image reconstruction process.}
\end{figure}

At the camera plane, the forward and backward generated signal beams, also known as object and reference beams, respectively, produce a single-photon hologram which intensity distribution can be calculated as
\begin{equation}
    I(\boldsymbol{r})\sim 1 + \cos^{2}{(\theta)}+2 T^{2}(\boldsymbol{r})\cos{(\theta)}\sin{\left[ \phi + \phi_{\mathrm{tilt}}(\boldsymbol{k}\cdot \boldsymbol{r}) + 2 \phi_{\mathrm{object}}(\boldsymbol{r})\right]},
\end{equation}

where $\theta$ is the relative angle between object and reference beams, $\phi$ is a global interferometric phase controlled experimentally, $\phi_{\mathrm{tilt}}(\boldsymbol{k}\cdot \boldsymbol{r})$ is a linear phase that depends on the relative angle between object and reference beams introduced by the mirror tip-tilt. $\phi_{\mathrm{object}}(\boldsymbol{r})$ is the spatially dependent phase and $T(\boldsymbol{r})$ is the spatially dependent amplitude transmission function of the object, the square and the factor of 2 counts for the double-pass through the object \cite{Fuenzalida2, Kundu, Lahiri2015, Marta}.


In the present scheme, momentum correlations are exploited \cite{Walborn2010, Grayson1994}, therefore the lenses L1 - L5 and L7 in Fig. 2 are placed such that the Fourier Plane (FP) of the crystal is projected onto the object plane, rightmost mirror and the BS. The lens L6 images the projected FP of the crystal onto the camera plane. In such a manner, we make use of the momentum correlations shared by the photon-pair to reconstruct an image by measuring a photon that never interacted with the object.

\section{Methods and results}
\subsection{Image reconstruction road map}

Figure 3 shows the images obtained at the different steps of the reconstruction process, this process is based on the classical OAH implementation \cite{Sanchez, Besaga2019, Cuche, Verrier}. 

First, the hologram, Eq. (3), is recorded on a sCMOS camera of $6.5$ $\mu$m pixel size, with 6 s exposure time and no gain, the field of view has a diameter of $11.9\pm 0.1$ mm. The raw image shows the single-photon hologram. Due to the insertion of the relative angle between object and reference beams, the fringe pattern displays a high spatial modulation shown in the light-blue highlighted region in which a small distortion of the fringes can be seen due to the presence of the phase object. Second, the raw image is treated in the FS by taking its fast Fourier transform (FFT). A zoom-in of the yellow highlighted region reveals the $+1$, DC and $-1$ terms characteristic of the Fourier OAH implementation. Third, the direct term ($+1$), enclosed in the dash black circle, is cropped and centered, maintaining the original image dimension (2048 x 2048), this region undergoes an IFFT which output is a complex-value image, which modulus is associated to the amplitude and the argument to the phase of the optical field that probes the sample, resulting in the amplitude and phase images \cite{Sanchez, Verrier}. A reference image of a glass plate without object features is previously recorded, its phase image is subtracted from the one containing the object features as the calibration of the system. This image is phase-unwrapped to obtain a flat and homogeneous phase distribution shown as the final image \cite{Kim, Cuche}.\\
\begin{figure}[ht!]
\centering\includegraphics[width=13.2cm]{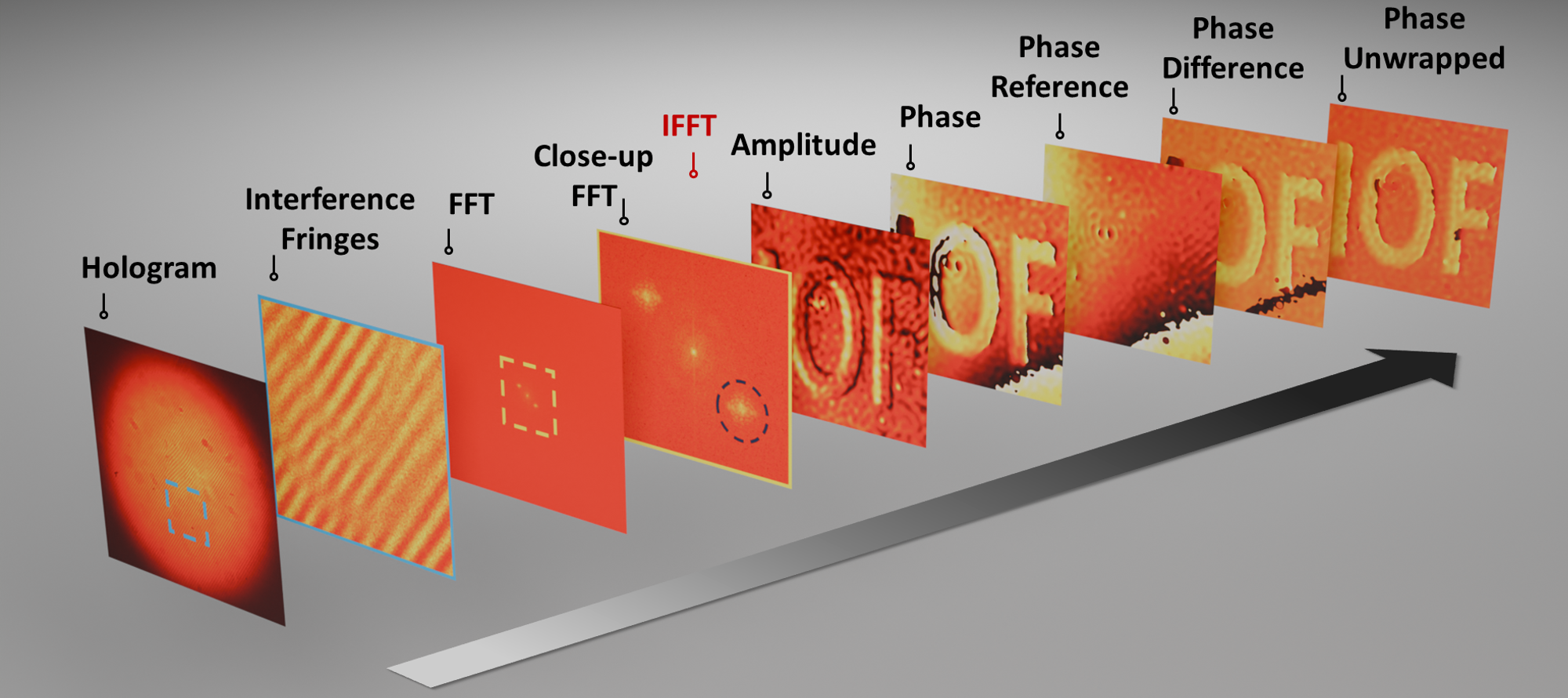}
\caption{Image reconstruction process, from left to right. Single-photon hologram recorded at the camera plane, zoom-in of the blue highlighted area containing the high density of fringes characteristic of OAH schemes. A FFT is performed to the hologram, a close-up of the yellow dashed square reveals the $+1$ term, in the black dashed circle. The direct terms is filtered and centered, maintaining the image dimensions. An IFFT is performed, amplitude and phase images are extracted. A reference phase image is taken and subtracted from the one containing the object features, to later be phase-unwrapped to obtain the final phase image.}
\end{figure}

\subsection{Imaging capabilities of QOAHUL}

In order to verify the imaging capabilities of the optical system we study the phase contrast while varying the exposure time, the influence of the spatial density of fringes on the overall image quality and the spatial resolution. These measurements are essential in imaging schemes to validate the performance, optimize parameters, ensure reliable image reconstruction and to asses the impact of artifacts.

Different objects are employed for the characterization of the setup. First, we use the standard binary object of the 1951 United States Air Force (USAF) resolution target, see Fig. 4 a). Its amplitude image is shown in Fig. 4 b). 

A second object is a glass plate with a miniaturized version of the USAF resolution target features engraved on it. The reconstructed phase image is presented in Fig. 4 c). The image is taken with 9 s exposure time. 

Additionally, a glass plate with engraved letters IOF is used. Its reconstructed phase image is presented in Fig. 4 d), an exposure time of 6 s is used in this case. Both phase objects with refractive index 1.6 are used to validate the efficacy of our technique, an estimation of the physical height of the engraved features corresponding to the while lines in Fig. 4 c) and d) are shown in Fig. 4 e) and f), respectively. The measurements are in good agreement with the manufacturer data $189 \pm 7$ nm for the USAF object, and $182 \pm 1$ nm for the IOF object. 

Finally, the system resolution allows to identify some features of a fly-wing, Fig. 4 g) shows its amplitude image. 

\begin{figure}[ht!]
\centering\includegraphics[width=13.2cm]{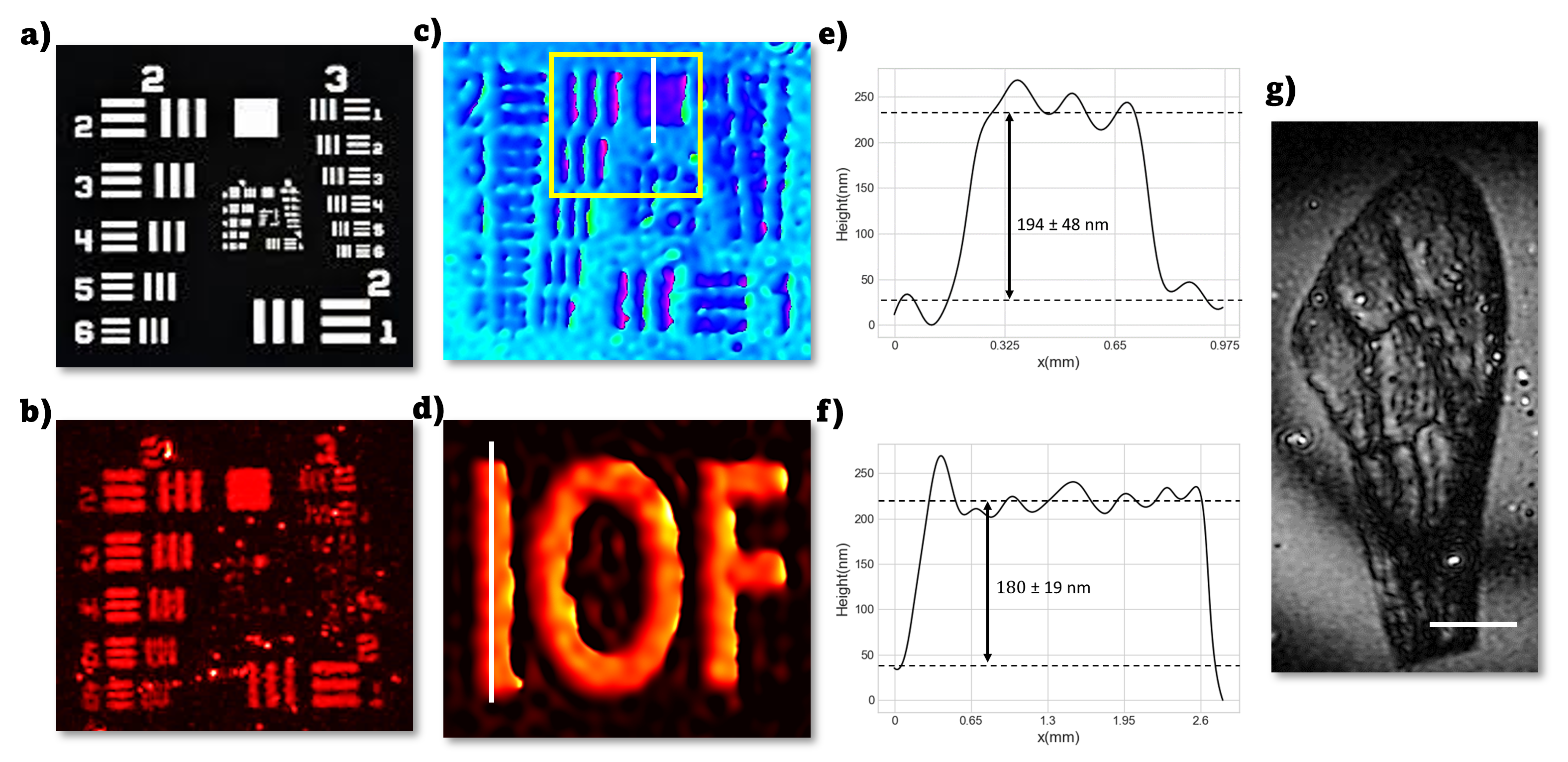}
\caption{Imaging capabilities of QOAHUL, a) image of the standard 1951 United States Air Force (USAF) resolution target, b) amplitude image of the USAF target. c) phase image of the USAF phase object, d) phase image of the glass plate with engraved IOF letters. e) and f) height profile of the USAF and IOF phase objects taken corresponding to the features with white lines of c) and d) respectively. g) Amplitude image of a fly-wing, the white bar has a length of 1.6 mm.}
\end{figure}

\subsubsection{Phase contrast vs exposure time}

In order to find the minimum exposure time in which the optical system is capable to retrieve the phase contrast of the USAF phase object we gradually vary the exposure time from 0.5 s to 9 s. The phase contrast is defined as the difference between the average phase values of the height and ground level features.

The upper row of Fig. 5 a) shows the amplitude images of the USAF phase object, and the lower row shows the phase images corresponding to the region highlighted by the yellow square in Fig. 4 c). The region enclosed by the black dashed rectangle in the lower row of Fig. 5 a) is utilized for the phase contrast analysis.

The longer the exposure time the better the ability of the system to reconstruct the object. After 1 s the features of the object are clearly distinguished. A strong agreement of our measurements and the expected phase contrast value is seen in Fig. 5 b), the data points from 1 s to 9 s are found around 1.95 radians (phase contrast given by manufacturer). A big deviation is found for 0.5 s integration time, the system is unable to reconstruct the object. Due to low signal to noise ratio (SNR) resulting images have more noise, which isn't suppressed well enough with the present signal. Those fluctuations in the phase values lead to scrambled images, as shown in Fig. 5 a) for 0.5 s exposure time. The SNR and the integration time could be easily improved by using a more powerful laser or/and a longer crystal \cite{Kutas2022}, in our case the pump power is of 20 mW and the crystal length of 2 mm, resulting in an estimate of $3.7 \cdot 10^{6}$ photons/mW/s reaching the camera plane. 

\begin{figure}[ht!]
\centering\includegraphics[width=13.2cm]{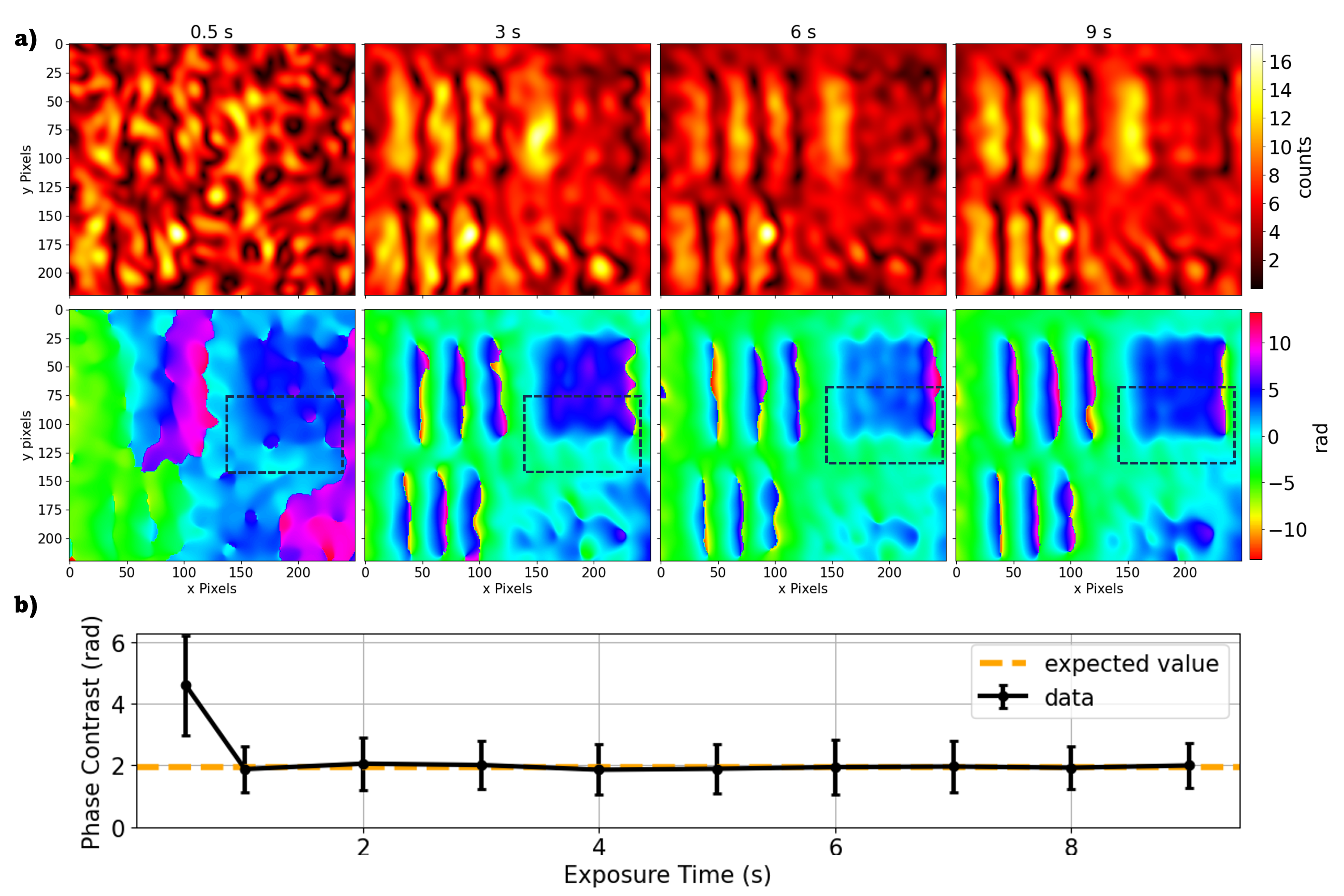}
\caption{Phase contrast vs exposure time. a) The amplitude and phase images of the USAF phase object are shown in the upper and lower rows, respectively. From left to right, the images are taken at 0.5, 3, 6 and 9 seconds exposure time, a gradual improvement on the image quality can be seen for longer exposure times. b) The exposure time is varied from 0.5 to 9 seconds. From 1 to 9 seconds the phase contrast is around the expected value of 1.95 radians given by the manufacturer. The regions enclosed by the black dashed rectangle are used in all the images for the analysis of the phase contrast.}
\end{figure}


\subsubsection{Phase contrast vs spatial period of the fringe pattern}

The impact of the relative angle between the object and the reference beams on the phase contrast and overall image quality is studied. This angle influences the opening of the $+1$ and $-1$ terms in the FS, Eq. (2). When the angle is too narrow, it hinders the ability to filter and isolate the direct term compromising the image quality by adding spatial frequency components of the DC term. Therefore, a wider angle guarantees an effortless filtering and isolation of the direct term, but its value has a classical upper bound given by the pixel size of the camera, $\theta_{max}=\sin^{-1}{(\lambda/2 d)}$. Assuming the small angle approximation (if $\lambda\ll d$), the maximum angle is $\theta_{max}\sim\lambda/2 d=4^{\circ}$, where $\lambda$ is the detected wavelength (910 nm) and $d$ is the pixel size (6.5 $\mu$m). The angle estimation is susceptible to errors due to the broad bandwidth of the SPDC spectra, but it serves as a reference.

\begin{figure}[ht!]
\centering\includegraphics[width=13.2cm]{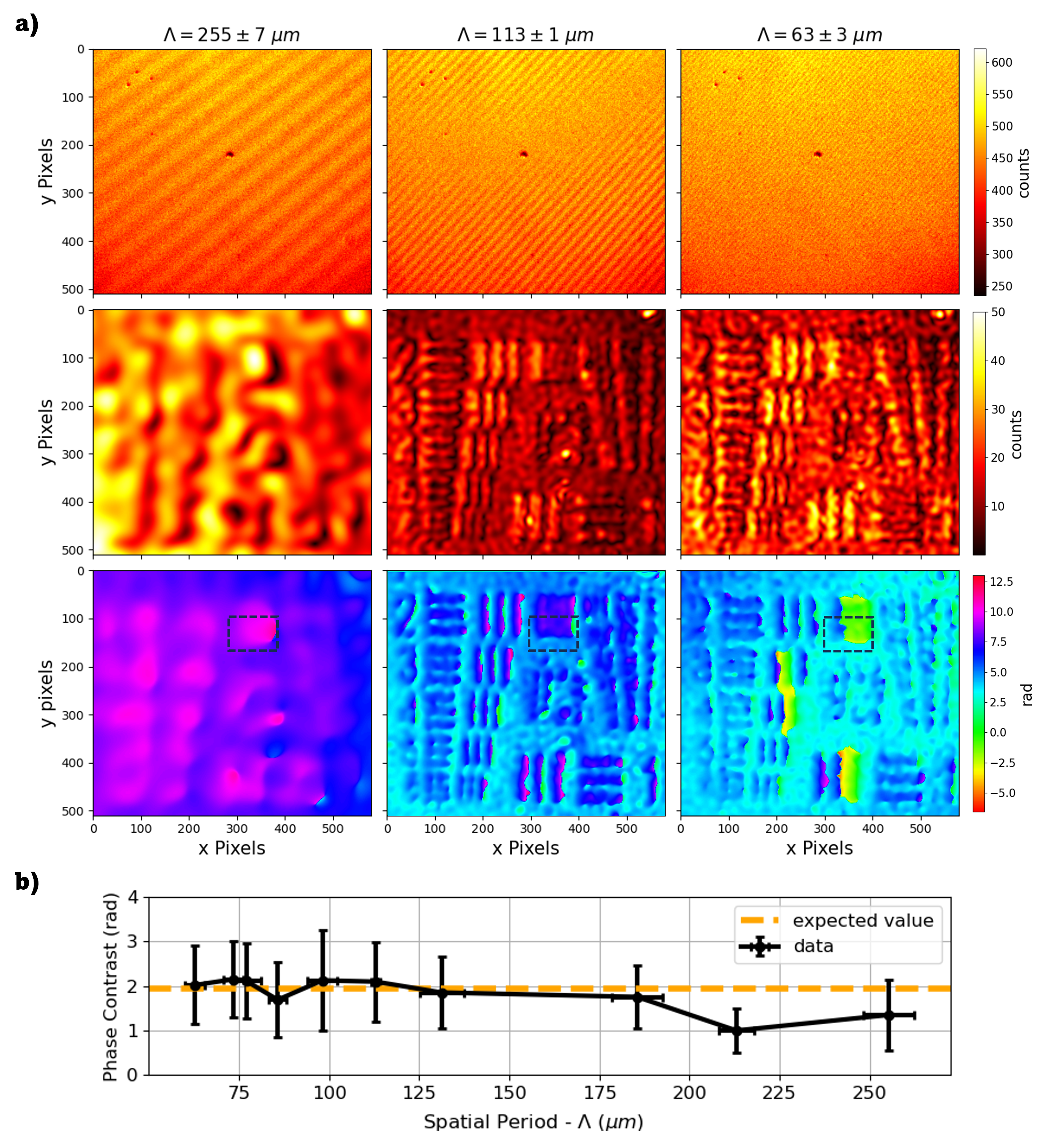}
\caption{Influence of the fringe pattern spatial period on the amplitude and phase images of the USAF phase object. a) Upper row shows a close up of the single-photon holograms, bright and dark fringes are visible, the spatial periods shortens from left to right. Middle and lower rows show the amplitude and phase images, respectively. A trade-off between the contrast and number of fringes is met for $\Lambda=113 \pm 1$ $\mu$m. b) The fringe pattern spatial period is varied from $63$ $ \mu$m to $255$ $ \mu$m, for short spatial periods the measurements are in good agreement with the phase contrast expected value. The regions enclosed by the black dashed rectangle are used in all the images for the analysis of the phase contrast.}
\end{figure}


This relative angle is complex to measure due to the spatial distribution of the SPDC spectra, therefore a practical quantity to examine this angle is the spatial period of the fringe pattern. Figure 6 shows the influence of the fringe pattern spatial period on the image quality of the optical system. The images are taken with 9 s exposure time. 

First, a close up of the single-photon holograms are shown in the upper row of Fig. 6 a), from left to right, the spatial period of dark and bright fringes take the following values: $\Lambda = 255\pm 7 $ $\mu$m , $\Lambda = 113 \pm 1$ $\mu$m, and $\Lambda = 63 \pm 3$ $\mu$m. 

A long spatial period hinders the ability of the system to retrieve the object information, even though it may posses a relative good fringe visibility, see first column of Fig. 6 a). A long spatial period equals a small relative angle for which the direct and conjugate terms do not separate far enough from the DC component, leading to less spatial frequency components of the object and thus poor image quality, see Sec. 2.1 and Fig. 1 c).

On the other hand, a very short spatial period does not guarantee the best performance due to its low fringe visibility, as shown in the third column of Fig. 6 a). In this scenario, the shorter the spatial period of the fringes the larger the relative angle between sample and reference beams causing the fringe visibility to drop as a consequence of losing the path-indistinguishability.

In Fig. 6 b) the effect of the spatial period on the phase contrast is analyzed. The largest deviation occurs at longer spatial periods, when the $+1$ and DC terms overlapped in the FS and a suitable reconstruction is not longer viable. From the graph, a weak impact of the fringe spatial period on the phase contrast can be inferred.


\subsubsection{Spatial resolution}

The spatial resolution of the optical system is characterized through the modulation transfer function (MTF), $MTF=\frac{\pi}{4}*CTF$, where $CTF$ is the contrast transfer function given by $CTF=\frac{I_{max}-I_{min}}{I_{max}+I_{min}}$ \cite{Coltman, Moreau}. Figure 7 shows the MTF values extracted from Fig. 4 b) and c) corresponding to the USAF amplitude object and the USAF phase object, respectively. The images are taken with 9 s exposure time and with a fringe pattern spatial period of $\Lambda=113 \pm 1$ $\mu$m.

We consider a line pair as resolvable when its MTF values for the horizontal and vertical bar patterns (including their uncertainty bars) are above 0.09 for amplitude objects and 0.27 for phase objects. Following the Rayleigh criterion \cite{Viswanathan2021}, the optical system is able to resolve a spatial frequency of 5.6 (lines/mm) which corresponds to a feature size of $89$ $\mu$m for both amplitude and phase objects. 

For the USAF amplitude object all the elements until the second element of group 3 with horizontal bar patterns are resolvable. For the USAF phase object all the elements with vertical bar patterns are resolvable, corresponding to a spatial frequency of 8.8 (lines/mm) and a feature size of $56.8$ $\mu$m. These differences between the MTF values obtained for the horizontal and vertical oriented bar patterns might be introduced by a slightly tilt in the object plane and a shift to its nominal position along the optical axis. 
 
\begin{figure}[ht!]
\centering\includegraphics[width=13.2cm]{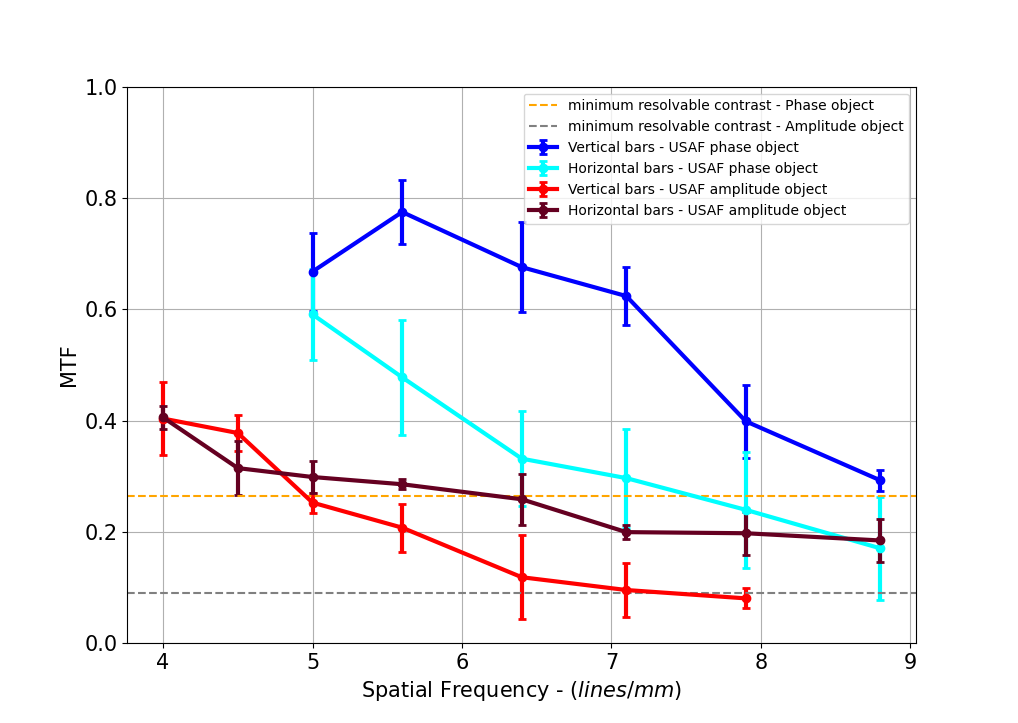}
\caption{ The USAF resolution target mask and phase objects are utilized to asses the spatial resolution of the optical system. Figure 4 b) and c) are analyzed to calculate the MTF of the system. Elements of the USAF target, vertical and horizontal bar patterns, for which the MTF values (including their errors bars) are above the dashed horizontal lines are considered resolvable.}
\end{figure}

\section{Discussion}

We present a novel imaging method that combines a hybrid-type induced-coherence non-linear interferometer and the principles of classical OAH into a new imaging scheme, QOAHUL. This technique retrieves amplitude and phase information of an object by exploiting the quantum correlations of a photon-pair generated by a non-linear crystal via SPDC. Our technique takes a major step forward respect to previous implementations \cite{Marta, Kviatkovsky2020, Kutas2022} by acquiring the full object information with a single measurement opening the possibility for the analysis of dynamic systems.

Our approach relies on the implementation of a hybrid-type induced-coherence non-linear interferometer for OAH, in which the signal beams propagate through a Mach-Zender interferometer while the idler beams through a Michelson-Morley interferometer \cite{Kim2024}. This provides a better control of the signal beams in comparison to the bidirectional approach, in which the signal beams recombine at the crystal plane after being back reflected by a mirror, rather than at the beam spliter, omitting the Mach-Zender interferometer. This imposes strong restrictions on the relative angle between object and reference beams and decreases the contrast of the fringes due to the lost of indistinguishability, bringing down the overall performance of the system. We would like to mention that during the writing of this manuscript we came across a parallel work showcasing the bidirectional approach \cite{Pearce}.

Our results represent a key step for the field of holography and contributes to the further development of holographic techniques based on non-linear interferometers with induced-coherence, representing a leap forward in fields such as quantitative phase imaging and phase contrast microscopy.

\begin{backmatter}
\bmsection{Funding}

This work was funded by the Deutsche Forschungsgemeinschaft (DFG, German Research Foundation) under Germany´s Excellence Strategy – EXC 2051 – Project-ID 390713860. 

In addition, this work was supported by the Horizon WIDERA 2021-ACCESS-03-01 gran 101079355 "BioQantSense" and from the European Union’s Horizon 2020 Research and Innovation Action under Grant Agreement No. 101113901 (Qu-Test, HORIZON-CL4-2022-QUANTUM-05-SGA).

We also acknowledge funding from the German Ministry of Education and Research (FKZ 13N14877, 13N16441), European Union’s Horizon 2020 research and innovation programme (Grant Agreement No. 899580, FastGhost) and from the FTI-Thüringen TECHNOLOGIE program, project 2020FGI0023.

\bmsection{Acknowledgments}

We would like to show our gratitude to C. Sevilla, J. Fuenzalida, S. Kodgirwar, R. Sondenheimer, P. Hendra, and V. Besaga for sharing their time and wisdom with J.R.L.T. during the course of this research. Furthermore, we thank to M. Safari Arabi, B. A. Matarrita Carranza and P. Snehrashmi Mehta for helping with the fly-wing sample preparation.

\bmsection{Author contributions} 

F.K., J.R.L.T., and M.K. conceived the idea. J.R.L.T., M.G.B., F.K., and M.K. designed the experiments and J.R.L.T. and F.K. performed the measurements. J.R.L.T. analyzed the data and wrote the paper. F.S., V.G., B.J., and M.G. supervised the project. All authors discussed the results.

\bmsection{Disclosures}

The authors declare no conflicts of interest.

\end{backmatter}

\bibliography{Optica-template}

\end{document}